\documentstyle[12pt]{article}

\newcommand{\om}{\omega}
\newcommand{\al}{\alpha}
\newcommand{\prt}{\partial}
\newcommand{\bt}{\beta}
\newcommand{\dlt}{\delta}
\newcommand{\gm}{\gamma}
\newcommand{\Gm}{\Gamma}
\newcommand{\Om}{\Omega}
\newcommand{\dgr}{\dagger}

\begin{document}

\begin{center}
{\large{\bf Interplay between Quantum and Coherent Effects in Optics} \\ [5mm]
V.I. Yukalov} \\ [2mm]
{\it Bogolubov Laboratory of Theoretical Physics\\
Joint Institute for Nuclear Research, Dubna 141980, Russia}

\end{center}

\vskip 1cm

\begin{abstract}

A technique is presented for treating strongly nonstationary and transient
processes in optics, permitting one to take into account both types of 
competing with each other effects, quantum as well as coherent. The main
equations for describing the interplay between these two kinds of effects
are derived. The possibility of influencing coherent optical phenomena by
preparing special quantum states of matter is discussed.

\end{abstract}

\vskip 1.5cm


\vskip  3cm

{\large {\bf 1. Introduction}}

\vskip 3mm

Coherent and quantum phenomena are often treated as antagonists, since 
the former are, to some extent, close to classical ones{\large$^1$}. When 
describing one of these phenomena, one usually employs rather different 
approximations. For example, a very common way of considering coherent phenomena 
is by invoking the semiclassical approximation, which makes it possible to give 
a transparent picture of such phenomena. But the semiclassical approximation 
kills all quantum correlations, eliminating by this all quantum effects. Such 
a neglection of the latter may be admissible if the studied coherent process is 
stationary or a strong coherence is imposed on the system by an external field. 
However, if one investigates a self-organized growth of coherence and when the 
coherent phenomena are transient or intermittent, then quantum effects may
essentially influence the features of coherent phenomena. In that case, it is 
necessary to take account of both types of effects. For this purpose, one may 
resort to the consideration of higher-order correlation functions. But then, the
evolution equations become so much complicated that essentially nonstationary 
states are hardly treatable.

In this report, a method is presented, which combines the transparency of the
semiclassical approximation with the possibility of taking account of quantum 
effects. The idea of the method reminds, to some extent, quantization of soliton
solutions in quantum field theory{\large$^2$}. But the most important concept 
of the method is the notice that quantum and coherent effects occur on rather 
different scales. Quantum effects are usually connected with short-range fast
fluctuations, while coherent phenomena are mainly long-range in space and slow
in time, as compared to the quantum ones. The occurrence of different 
spatio-temporal scales allows the development of {\it Scale Separation 
Approach}{\large$^{3-7}$}. The general idea of separating different scales 
is, of course, known. However, its realization, as applied to optics in the 
report below is new. The pivotal novel technique is {\it Quantization of 
Local Fluctuations}.

\vskip 5mm

{\large {\bf 2. Resonant Atoms in Matter}}

\vskip 3mm

The system of resonant atoms inside matter is described by the Hamiltonian
\begin{equation}
\label{1}
\hat H = \hat H_a + \hat H_f + \hat H_m + \hat H_{af} + \hat H_{mf} \; .
\end{equation}
Here, the first term
\begin{equation}
\label{2}
\hat H_a = \sum_{i=1}^N \om_0\left ( S_i^z +\frac{1}{2}\right )
\end{equation}
is the Hamiltonian of resonant atoms, with transition frequency $\om_0$, where
$S_i^\al$ are pseudospin operators. The field Hamiltonian is
\begin{equation}
\label{3}
\hat H_f = \frac{1}{8\pi} \int \left ( {\bf E}^2 + {\bf B}^2 \right ) \;
d{\bf r} \; .
\end{equation}
The Gaussian system of units is used, with setting $\hbar\equiv 1$. Electric
and magnetic fields are expressed through the vector potential,
$$
{\bf E}= -\;\frac{1}{c}\; \frac{\prt {\bf A}}{\prt t} \; , \qquad
{\bf B} = {\bf\nabla}\times {\bf A} \; .
$$
The Coulomb gauge ${\bf\nabla}\cdot{\bf A}=0$ will be employed. This calibration 
is convenient for eliminating field variables in the following evolution
operations. The Hamiltonian $\hat H_m$ models the material incorporating the
atoms. The atom-field interaction is presented by the dipole Hamiltonian
\begin{equation}
\label{4}
\hat H_{af} = - \sum_{n=1}^N \left ( \frac{1}{c}\; {\bf J}_i\cdot {\bf A}_i +
{\bf P}_i \cdot {\bf E}_{0i} \right ) \; ,
\end{equation}
in which ${\bf A}_i\equiv{\bf A}({\bf r}_i,t)$ is the vector-potential operator,
${\bf E}_{0i}\equiv {\bf E}_0({\bf r}_i,t)$ is a classical external field. The
transition current and transition dipole operators are
\begin{equation}
\label{5}
{\bf J}_i = i\om_0 \left ( {\bf d} S_i^+ - {\bf d}^* S_i^- \right ) \; , 
\qquad {\bf P}_i = {\bf d} S_i^+ + {\bf d}^* S_i^- \; ,
\end{equation}
where ${\bf d}$ is a transition dipole and $S_i^\pm \equiv S_i^x\pm iS_i^y$.
Finally, the matter-field interaction is described by the Hamiltonian
\begin{equation}
\label{6}
\hat H_{mf} = -\; \frac{1}{c} \int {\bf j}_m({\bf r},t) \cdot
{\bf A}({\bf r},t)\; d{\bf r}\; ,
\end{equation}
in which ${\bf j}_m({\bf r},t)$ is a local density of current in matter.

The Heisenberg equations of motion for the pseudospin operators yield the 
equation 
\begin{equation}
\label{7}
\frac{d S_i^-}{dt} = - i\om_0 S_i^- + 2 \left ( k_0 {\bf d}\cdot {\bf A}_i -
i{\bf d}\cdot{\bf E}_{0i}\right ) S_i^z
\end{equation}
for the lowering operator and the Hermitian conjugated equation for the rising 
operator, with $k_0\equiv \om_0/c$. The equation for the population-difference
operator is
\begin{equation}
\label{8}
\frac{dS_i^z}{dt} = - \left ( k_0{\bf d}\cdot {\bf A}_i - i {\bf d}\cdot
{\bf E}_{0i}\right ) S_i^+ - \left ( k_0{\bf d}^*\cdot{\bf A}_i +
i{\bf d}^*\cdot{\bf E}_{0i}\right ) S_i^- \; .
\end{equation}
In deriving an equation for the vector potential, one has to use the commutation
relations
$$
\left [ E^\al({\bf r},t),\; A^\bt({\bf r}',t)\right ] = 4 \pi ic\dlt_{\al\bt}
({\bf r}-{\bf r}') \; ,
$$
where
$$
\dlt_{\al\bt} ({\bf r}) \equiv \frac{1}{(2\pi)^3} \int \left ( \dlt_{\al\bt} 
-\; \frac{k^\al k^\bt}{k^2}\right ) e^{i{\bf k}\cdot{\bf r}}\; d{\bf k}
$$
is the so-called transverse delta-function{\large$^1$}. These commutation 
relations explicitly take into account the Coulomb gauge condition. The
transverse delta-function can be presented in other forms, as
$$
\dlt_{\al\bt}({\bf r}) = \dlt_{\al\bt}\dlt({\bf r}) +
\frac{\prt^2}{\prt r^\al\prt r^\bt} \int 
\frac{e^{i{\bf k}\cdot{\bf r}}}{(2\pi)^3 k^2}\; d{\bf k} \; ,
$$
or as
$$
\dlt_{\al\bt}({\bf r}) = \frac{2}{3}\; \dlt_{\al\bt}\;
\dlt({\bf r}) - \; \frac{\dlt_{\al\bt} -3n_\al n_\bt}{4\pi r^3} \; ,
$$
where $n_\al\equiv r^\al/r$. As a result, for the vector potential one has
\begin{equation}
\label{9}
\left ( {\bf\nabla}^2 -\; \frac{1}{c^2}\; \frac{\prt^2}{\prt t^2}\right )
{\bf A} = -\; \frac{4\pi}{c}\; {\bf j} \; ,
\end{equation}
with the density of current
\begin{equation}
\label{10}
j^\al({\bf r},t) = \sum_\bt \left [ 
\sum_{i=1}^N \dlt_{\al\bt}({\bf r}-{\bf r}_i') J_i^\bt (t) + 
\int \dlt_{\al\bt}({\bf r}-{\bf r}') j_m^\bt({\bf r}',t)\; 
d{\bf r}' \right ] \; .
\end{equation}

The solution to Eq. (9) reads
\begin{equation}
\label{11}
{\bf A}({\bf r},t) =  {\bf A}_{vac}({\bf r},t) + \frac{1}{c} \int
{\bf j}\left ({\bf r}',t-\; \frac{|{\bf r}-{\bf r}'|}{c}\right ) \; 
\frac{d{\bf r}'}{|{\bf r}-{\bf r}'|} \; ,
\end{equation}
with ${\bf A}_{vac}$ being the solution of the uniform equation related to 
Eq. (9). Note that the dipolar part of the transverse delta-function, being 
averaged over space, yileds zero, that is
$$
\int \dlt_{\al\bt}({\bf r})\; d{\bf r} = \frac{2}{3}\; \dlt_{\al\bt} \; .
$$
This anisotropic dipolar part, for simplicity, can be omitted. Then the vector
potential (11) takes the form
\begin{equation}
\label{12}
{\bf A} = {\bf A}_{vac} + {\bf A}_{rad} + {\bf A}_{mat} \; ,
\end{equation}
in which the term
$$
{\bf A}_{rad}({\bf r}_i,t) = \sum_{j(\neq i)}^N \frac{2}{3c r_{ij}}\;
{\bf J}_j\left ( t -\; \frac{r_{ij}}{c}\right )
$$
is due to radiating atoms and the potential
$$
{\bf A}_{mat}({\bf r},t) = \frac{2}{3c} \int {\bf j}_m 
\left ( {\bf r}',t -\; \frac{1}{c} \; |{\bf r} - {\bf r}'|\right ) \;
\frac{d{\bf r}'}{|{\bf r}-{\bf r}'|}
$$
is caused by matter currents. Here $r_{ij}\equiv|{\bf r}_{ij}|$, 
${\bf r}_{ij}\equiv{\bf r}_i -{\bf r}_j$. The summation over $j=1,2,\ldots,N$ 
does not include the term with $j=i$ corresponding to self-action, which will
be taken into account by incorporating the level and line widths in the 
evolution equations.

It is important to notice that all processes are defined for $t\geq 0$, while 
in the expressions above there appears the dependence of operators on the
difference $t-t'$, which may be negative. Therefore it is necessary to
complete the definition of quasispin operators by adding the retardation 
condition
\begin{equation}
\label{13}
S_j^-(t)= 0 \qquad (t< 0) \; .
\end{equation}
The dependence of $S_j^-(t-t')$ on the retarded time can also be simplified 
by taking into account that the interaction of radiation with atoms is to 
be essentially weaker than intra-atomic interactions. In the other case,
the very notion of resonant atoms, having a well-defined transition frequency,
would not have sense. Then, Eq. (7) shows that $S_j^-(t)\sim \exp(-i\om_0 t)$.
Combining this with the retardation condition (13) gives 
\begin{equation}
\label{14}
S_j^-(t-t') =\Theta(t-t') S_j^-(t)\exp(i\om_0 t') \; .
\end{equation}
Hence for the typical dependence on the retarded variables, one has
\begin{equation}
\label{15}
S_j^-\left ( t -\; \frac{r_{ij}}{c}\right ) = \Theta(ct-r_{ij})
S_j^-(t)\exp(ik_0r_{ij}) \; .
\end{equation}
Such a simplification, combining the Born approximation with the retardation 
condition (13), can be called the {\it Retarded Born Approximation}.

It is worth mentioning that it is sufficient to understand all operator 
equations in the weak sense, as the equations for appropriate matrix 
elements or averages.

\vskip 5mm

{\large {\bf 3. Separation of Local Fields}}

\vskip 3mm

The influence of vacuum and matter on an atom enters the evolution equations
through the expression
\begin{equation}
\label{16}
\xi({\bf r},t) \equiv 2k_0{\bf d}\cdot\left ( {\bf A}_{vac} +
{\bf A}_{mat}\right ) \; .
\end{equation}
Actually, electromagnetic vacuum and matter form an effective vacuum acting
on an atom by means of the local field (16). Thus, one may say that there are
two types of operator variables, the quasispin variables and the local field 
(16). An operator $\hat F$, being a function of these variables, can, for
brevity, be written as $\hat F(S,\xi)$, with $S$ denoting the quasispin set 
$\{ S_i^\al\}$ and $\xi$, the set $\{\xi({\bf r},t)\}$ of local fields at 
different points ${\bf r}$. According to the existence of two types of 
variables, one may define two kinds of statistical averages over the related
variables, either quasispin or local-field ones. Such partial averages are 
defined by means of the corresponding restricted traces, either
\begin{equation}
\label{17}
<\hat F>\; \equiv {\rm Tr}_S\; \hat\rho\hat F(S,\xi) \; ,
\end{equation}
with $\hat\rho$ being a statistical operator, or
\begin{equation}
\label{18}
\ll \hat F\gg\; \equiv {\rm Tr}_\xi\; \hat\rho \hat F(S,\xi) \; .
\end{equation}

When averaging over quasispin variables, one may assume the validity of the
mean-field decoupling
\begin{equation}
\label{19}
<S_i^\al S_j^\bt>\; = \; < S_i^\al><S_j^\bt> \qquad (i\neq j) \; ,
\end{equation}
since atoms interact with each other through effective long-range forces.
This decoupling is not equivalent to the semiclassical approximation as the
local-field variables have not been involved. Then the atomic average
$<S_i^\al>$ is, actually, an operator function of local fields. The latter 
describe local quantum fluctuations.

For what follows, it is convenient to use the notation 
$S_i^\al(t)\equiv S^\al({\bf r}_i,t)$. Also, the summation over atoms can be 
replaced by the integration over the sample, according to the rule
\begin{equation}
\label{20}
\sum_{i=1}^N \Longrightarrow  \rho \int d{\bf r} \qquad \left ( \rho
\equiv \frac{N}{V} \right ) \; .
\end{equation}

The behaviour of atoms can be described by the following averages. For an 
atom at the point ${\bf r}$, one may write a transition function
\begin{equation}
\label{21}
u({\bf r},t) \equiv 2 <S^-({\bf r},t)>\; ,
\end{equation}
intensity of coherence
\begin{equation}
\label{22}
w({\bf r},t) \equiv u^+({\bf r},t) u({\bf r},t) \; ,
\end{equation}
and population difference
\begin{equation}
\label{23}
s({\bf r},t) \equiv 2<S^z({\bf r},t)> \; .
\end{equation}

The effective force acting on an atom is given by the sum
\begin{equation}
\label{24}
f({\bf r},t) = f_0({\bf r},t) + f_{rad}({\bf r},t) + \xi({\bf r},t) \; ,
\end{equation}
in which
\begin{equation}
\label{25}
f_0({\bf r},t) \equiv -2i{\bf d}\cdot {\bf E}_0({\bf r},t)
\end{equation}
is due to a classical external field,
\begin{equation}
\label{26}
f_{rad}({\bf r},t) \equiv 2k_0\; <{\bf d}\cdot {\bf A}_{rad}({\bf r},t)>
\end{equation}
is caused by the radiation of other atoms, and $\xi({\bf r},t)$ is the local 
field (16). The radiation force (26) explicitly writes
\begin{equation}
\label{27}
f_{rad}({\bf r},t) = - i\gm_0\rho \int \left [ 
G({\bf r}-{\bf r}',t)\; u({\bf r}',t) - {\bf e}_d^2 \; G^*({\bf r}-{\bf r}',t)
\; u^+({\bf r}',t)\right ] \; d{\bf r}' \; ,
\end{equation}
where the transfer function is
$$
G({\bf r},t) \equiv \Theta(ct-r)\; \frac{\exp(ik_0r)}{k_0r}
$$
and the notation
$$
\gm_0 \equiv \frac{2}{3}\; k_0^3 d_0^2 \; , \qquad r\equiv |{\bf r}|\; , 
\qquad {\bf d}\equiv d_0{\bf e}_d\; , \qquad d_0\equiv |{\bf d}|
$$
is used. The quantity $\gm_0$ is a natural half-width.

In this way, for the functions $u=u({\bf r},t)$, $w=w({\bf r},t)$, and 
$s=s({\bf r},t)$, we derive the evolution equations
\begin{equation}
\label{28}
\frac{\prt u}{\prt t} = - (i\om_0 +\gm_2) u + f s \; ,
\end{equation}
\begin{equation}
\label{29}
\frac{\prt w}{\prt t} = - 2\gm_2 w + \left ( u^+ f + f^+ u\right ) s \; ,
\end{equation}
\begin{equation}
\label{30}
\frac{\prt s}{\prt t} = -\; \frac{1}{2}\left ( u^+ f + f^+ u \right ) -
\gm_1 (s  - \zeta) \; ,
\end{equation}
in which $\gm_1$ is a longitudinal relaxation parameter, $\gm_2$ is a 
transverse attenuation parameter, and $\zeta\in [-1,1]$ is a stationary
population difference for a single atom.

\vskip 5mm

{\large {\bf 4. Sample of Cylindrical Shape}}

\vskip 3mm

Let the sample have typical for lasers cylindrical shape. The axis of the
cylinder is along the $z$-axis, which is distinguished by the propagating
field
\begin{equation}
\label{31}
{\bf E}_0({\bf r},t) = \frac{1}{2}\; {\bf E}_1 e^{i(kz-\om t)} +
\frac{1}{2} \; {\bf E}^*_1 e^{-i(kz-\om t)} \; ,
\end{equation}
where the frequency $\om=kc$ is close to the transition frequency $\om_0$,
so that the resonance condition
\begin{equation}
\label{32}
\frac{|\Delta|}{\om_0} \ll 1 \qquad (\Delta\equiv\om -\om_0 )
\end{equation}
holds. The wavelength $\lambda=2\pi c/\om$ is small as compared to the 
radius, $R$, and length, $L$, of the cylinder, 
\begin{equation}
\label{33}
\frac{\lambda}{R} \ll 1 \; , \qquad \frac{\lambda}{L} \ll 1 \; .
\end{equation}
Assuming the absence of sharp transverse structures, one may employ the
single-mode approximation
\begin{equation}
\label{34}
u({\bf r},t) = u(t) e^{ikz} \; , \qquad w({\bf r},t) = w(t) \; , \qquad
s({\bf r},t) = s(t) \; .
\end{equation}
The absence of sharp transverse nonuniformity implies that there exists 
the main propagating mode selected by the field (31). Expressions (34)
are to be substituted in Eqs. (28) to (30). Equation (28) is multiplied
by $e^{-ikz}$ and then all equations are averaged over space. The following
notation will be used for an effective force
\begin{equation}
\label{35}
f_1(t) \equiv - i{\bf d}\cdot {\bf E}_1 e^{-i\om t} + \xi(t) \; ,
\end{equation}
in which
\begin{equation}
\label{36}
\xi(t) \equiv \frac{1}{V} \int \xi({\bf r},t) e^{-ikz}\; d{\bf r} \; .
\end{equation}
Let us also introduce the coupling functions
\begin{equation}
\label{37}
\al(t) \equiv \gm_0\rho \int\Theta(ct-r)\; \frac{\sin(k_0r -kz)}{k_0r}\; 
d{\bf r}\; , \quad
\bt(t) \equiv \gm_0\rho \int\Theta(ct-r)\; \frac{\cos(k_0r -kz)}{k_0r}\; 
d{\bf r}\; .
\end{equation}
Finally, Eqs. (28) to (30) are transformed into the ordinary differential
equations
\begin{equation}
\label{38}
\frac{du}{dt} = -\left [ i(\om_0 +\bt s) +\gm_2 - \al s\right ] u + f_1 s\; ,
\end{equation}
\begin{equation}
\label{39}
\frac{dw}{dt} = -2(\gm_2 -\al s) w + \left ( u^+ f_1 + f_1^+ u\right ) s\; ,
\end{equation}
\begin{equation}
\label{40}
\frac{ds}{dt} = -\al w -\; \frac{1}{2} \left ( u^+ f_1 + f_1^+ u\right ) -
\gm_1 (s - \zeta ) \; .
\end{equation}
Although, one should remember that these are, actually, operator equations 
with respect to the quantum variable $\xi(t)$.

\vskip 5mm

{\large {\bf 5. Method of Stochastic Averaging}}

\vskip 3mm

To simplify further the evolution equations (38) to (40), one can take into 
consideration the existence of several small parameters, such as
\begin{equation}
\label{41}
\frac{\gm_0}{\om_0} \ll 1 \; , \qquad \frac{\gm_1}{\om_0} \ll 1 \; , \qquad 
\frac{\gm_2}{\om_0} \ll 1 \; .
\end{equation}
The amplitude of the external field (31) is assumed to be small, so that
\begin{equation}
\label{42}
\frac{|\nu_1|}{\om_0} \ll 1 \; , \qquad 
\nu_1 \equiv {\bf d}\cdot {\bf E}_1 \; .
\end{equation}
Also, the local quantum field $\xi$ is treated as weak, in the sense that 
its first and second moments are proportional to values much smaller than 
$\om_0$. Then the multiscale averaging technique can be 
generalized{\large$^{3-7}$} to stochastic and operator equations, as Eqs. (38)
to (40). The occurrence of the above small parameters shows that the functions
$w(t)$ and $s(t)$ are temporal quasi-invariants with respect to the fast 
function $u(t)$.
				
Equation (38) for the fast function, with $w$ and $s$ being quasi-invariants,
can be solved. To this end, let us introduce the {\it collective width} and
{\it collective frequency} by the corresponding expressions
\begin{equation}
\label{43}
\Gm\equiv \gm_2 -\al s \; , \qquad \Om \equiv \om_0 +\bt s \; ,
\end{equation}
and also, let us define the {\it dynamical detuning}
\begin{equation}
\label{44}
\delta \equiv \om - \Om = \Delta -\bt s \; .
\end{equation}
The solution of Eq. (38) reads
\begin{equation}
\label{45}
u =\left ( u_0 -\; \frac{\nu_1 s}{\delta + i\Gamma}\right ) e^{-(i\Om+\Gm)t}
+ \frac{\nu_1 s}{\delta+i\Gm}\; e^{-i\om t} +
s \int_0^t \xi(t') e^{-(i\Om+\Gm)(t-t')}\; dt' \; .
\end{equation}
To simplify the following formulas, it is convenient to choose the phase of
${\bf E}_1$ so that to eliminate the dependence on the transverse initial 
value $u_0$. For this purpose, the phase of ${\bf E}_1$ is taken so that
$u_0^*{\bf d}\cdot{\bf E}_1$ be real, which writes
\begin{equation}
\label{46}
u_0^* \nu_1 = u_0 \nu_1^* \; .
\end{equation}

The solution (45) is to be substituted into Eqs. (39) and (40), whose
right-hand sides are to be averaged according to the prescription
$$
\ll \overline F \gg \; = \lim_{T\rightarrow\infty} \;
\frac{1}{T}\; \int_0^T \ll F(\xi,t)\gg \; dt \; ,
$$
where the integration over time does not touch quasi-invariants. The quantum
field $\xi(t)$ is centered so that
\begin{equation}
\label{47}
\ll \xi(t) \gg \; = 0 \; .
\end{equation}
To make the resulting expressions less cumbersome, let us consider the case 
of small detuning $|\delta|<|\Gm|$, when $\delta$ can be omitted in the
phase dependence, though should be kept in denominators to avoid spurious
poles. In the process of the averaging, one obtains the {\it effective 
attenuation}
\begin{equation}
\label{48}
\tilde\Gm \equiv \frac{|\nu_1|^2\Gm}{\delta^2+\Gm^2} \left (
1 - e^{-\Gm t}\right ) + \Gm_3 \; ,
\end{equation}
where the first term is due to the classical external field (31) and the
{\it quantum attenuation}
\begin{equation}
\label{49}
\Gm_3 \equiv {\rm Re}\; \lim_{T\rightarrow\infty} \;
\frac{1}{T} \int_0^T dt \; \int_0^t \ll \xi^+(t)\xi(t')\gg \;
e^{-(i\Om+\Gm)(t-t')}\; dt'
\end{equation}
appears because of the action of the quantum field $\xi(t)$.

Equations (39) and (40) reduce to the evolution equations
\begin{equation}
\label{50}
\frac{dw}{dt} = -2(\gm_2-\al s)w + 2\tilde\Gm s^2 \; , \qquad
\frac{ds}{dt} = -\al w -\tilde\Gm s -\gm_1(s-\zeta) \; ,
\end{equation}
describing the coherent guiding centers. In order to make these equations 
complete, it is necessary to define the quantum correlation function 
$\ll\xi^+(t)\xi(t')\gg$ entering the quantum attenuation (49).

\vskip 5mm

{\large {\bf 6. Examples of Quantum Attenuation}}

\vskip 3mm

To make it clear how the quantum attenuation (49) can be calculated, let us
give some examples of defining the quantum variable $\xi(t)$.

The first simple case could be by considering the variable $\xi(t)$ as
random, associated to infrared noise characterized by
\begin{equation}
\label{51}
\ll \xi^+(t)\xi(t')\gg \; = \gm_3^2 \; .
\end{equation}
Then Eq. (49) yields
\begin{equation}
\label{52}
\Gm_3 = \frac{\gm_3^2\Gm}{\Om^2+\Gm^2} \; .
\end{equation}

The opposite case would be to treat $\xi(t)$ as a stochastic variable 
representing white noise, with
\begin{equation}
\label{53}
\ll \xi^+(t)\xi(t')\gg \; = 2\Gm_3 \delta(t-t') \; ,
\end{equation}
which results in the identity $\Gm_3=\Gm_3$.

A more elaborate modelling of the effective quantum fluctuations is by a 
system of oscillators, yielding
\begin{equation}
\label{54}
\xi(t) = \sum_q \gm(\om_q) \left ( b_q e^{-i\om_q t} + 
b_q^\dgr e^{i\om_q t} \right ) \; ,
\end{equation}
where $\om_q=\om_{-q}>0$. For the Bose operators $b_q$ and $b_q^\dgr$,
statistical averaging gives
\begin{equation}
\label{55}
\ll b_q\gg\; = 0 \; , \qquad \ll b_q^\dgr b_p\gg \; = n_q\delta_{qp} \; ,
\qquad \ll b_q b_p^\dgr\gg \; = ( 1 + n_q ) \delta_{qp} \; ,
\end{equation}
with $n_q$ being a momentum distribution function. The averages 
$\ll b_q b_q\gg$ and $\ll b_q^\dgr b_q^\dgr \gg $ in the case of normal 
effective vacuum are zero, while for a squeezed vacuum
\begin{equation}
\label{56}
\ll b_q^\dgr b_p^\dgr \gg \; = m_q \Delta(\om_q + \om_p -2\om_s) \; ,
\end{equation}
where the function $m_q$ is defined by the particular properties of a squeezed
vacuum, and $\Delta(\om)$ is the discrete delta-function
\begin{eqnarray}
\nonumber
\Delta(\om) \equiv \left \{ \begin{array}{cc}
1\; , & \om=0 \\
0\; , & \om \neq 0 \; . \end{array} \right.
\end{eqnarray}
The spectral function $\gm(\om)$ in Eq. (54) is assumed to be symmetric 
with respect to the central line, so that
$$
\gm (\om_s + \om_q ) = \gm (\om_s - \om_q ) \; .
$$

Such an effective squeezed vacuum can be realized if atoms are inserted into
a medium with squeezed collective excitations interacting with electromagnetic 
field. These excitations could be squeezed optical phonons or magnons. The
interaction of the latter with photons results in the formation of squeezed 
polaritons.

For the quantum field (54), the correlation function is 
$$
\ll \xi^+(t)\xi(t')\gg \; = \sum_q \gm^2(\om_q) \left [
n_q e^{i\om_q(t-t')} + (1+n_q) e^{-i\om_q(t-t')} + \right.
$$
\begin{equation}
\label{57}
\left. + m_q e^{i\om_q t+i(2\om_s-\om_q)t'} + m_q^* e^{-i\om_q t - 
i(2\om_s -\om_q) t'}\right ] \; .
\end{equation}

Calculating the quantum attenuation (49), we keep in mind that 
$n_q\equiv n(\om_q)$ is real, while $m_q\equiv m(\om_q)$ is, in general,
complex,
$$
m(\om) = |m(\om)| e^{i\varphi_s} \; .
$$
The final formulas will be simplified by remembering that $|\delta|\equiv
|\om-\Om|\ll \om$. Then substituting the correlation function (57) into
Eq. (49), one finds
\begin{equation}
\label{58}
\Gm_3 = \Gm \sum_q \gm^2(\om_q) \left [ \frac{n_q}{(\Om-\om_q)^2+\Gm^2} +
\frac{1+n_q}{(\Om+\om_q)^2+\Gm^2}\right ] - n(\om)\;
\frac{\gm^2(\om)\Gm}{\delta^2+\Gm^2}\; e^{-\Gm t} + \Gm_s \; ,
\end{equation}
with the last term being due to the squeezing,
\begin{equation}
\label{59}
\Gm_s = - |m(\om)| \gm^2(\om) \;
\frac{\Gm\cos\varphi_s + 2\om_s\sin\varphi_s}{4\om_s^2+\Gm^2}\;
e^{-\Gm t} \; .
\end{equation}
The quantity (59) may be named the {\it squeezing attenuation}.

Separating out of the quantum attenuation (58) its resonant part, one has
\begin{equation}
\label{60}
\Gm_{res} = n(\om) \; \frac{\gm^2(\om)\Gm}{\delta^2+\Gm^2}\left (
1 - e^{-\Gm t}\right ) \; .
\end{equation}
As is seen, the {\it resonant attenuation} (60) is zero at $t=0$. This means
that, at the initial time, the main role in the quantum attenuation is played
by its nonresonant parts, including the squeezing attenuation.

To complete this section, it is useful to present explicit examples of 
$n(\om)$ and $|m(\om)|$ in the case of a squeezed effective vacuum. The latter
can be generated by a parametric oscillator{\large$^1$}, with a squeezing
field proportional to $\varepsilon\cos(2\om_s t +\varphi_s)$. One employs the
notation $\mu\equiv\gm-\varepsilon$ and $\nu\equiv\gm+\varepsilon$, where 
$\gm$ is a cavity damping rate. Then, for a non-degenerate parametric 
oscillator, one has
$$
n(\om) = \frac{\nu^2-\mu^2}{8}\left [ \frac{1}{(\Delta_s+\kappa)^2+\mu^2}
+ \frac{1}{(\Delta_s-\kappa)^2+\mu^2}\; - \; 
\frac{1}{(\Delta_s+\kappa)^2+\nu^2}\; - \; \frac{1}{(\Delta_s-\kappa)^2+\nu^2}
\right ] \; ,
$$
$$
|m(\om)| = \frac{\nu^2-\mu^2}{8}\left [ \frac{1}{(\Delta_s+\kappa)^2+\mu^2}
+ \frac{1}{(\Delta_s-\kappa)^2+\mu^2} + \frac{1}{(\Delta_s+\kappa)^2+\nu^2} 
+ \frac{1}{(\Delta_s-\kappa)^2+\nu^2} \right ] \; ,
$$
where $\Delta_s\equiv\om-\om_s$. The parameter $\kappa$ characterizes a 
two-mode squeezed field, representing the displacement from the central 
frequency of squeezing, where the two-mode squeezed vacuum is maximally 
squeezed. If the parametric oscillator is weakly nondegenerate, with 
$|\kappa|\ll |\Delta_s|$, or degenerate, then 
$$
n(\om) = \frac{\nu^2-\mu^2}{4}\left ( \frac{1}{\Delta_s^2+\mu^2}\;
- \; \frac{1}{\Delta_s^2+\nu^2} \right ) \; , \qquad
|m(\om)| = \frac{\nu^2-\mu^2}{4}\left ( \frac{1}{\Delta_s^2+\mu^2}\;
+ \; \frac{1}{\Delta_s^2+\nu^2} \right ) \; .
$$
By means of different parametric oscillators, it is possible to generate 
squeezed fields with various characteristics.

\vskip 5mm

{\large {\bf 7. Conclusion}}

\vskip 3mm

The method, presented in this report, makes it possible to combine the 
clarity of the semiclassical approximation with taking account of quantum 
effects. The latter are responsible for the appearance of a specific quantum
attenuation. The vacuum electromagnetic field interacting with matter forms 
an effective quantum vacuum. Preparing different kinds of such vacua, one 
can regulate the value of the quantum attenuation. The latter, in its turn, 
can essentially influence the features of the following coherent processes.

Notice that the squeezing attenuation (59) can change sign for the varying 
phase $\varphi_s$. Thus, $\Gm_s<0$ for $\varphi_s=0,\pi/2$.

For preparing different types of effective vacua, one may incorporate 
resonant atoms into different media.  There exists a rich variety of 
materials possessing various properties, which could be used for creating 
effective  vacua with divers specific features. Just as a few cases of 
materials with interesting electromagnetic properties, one may mention 
biopolymers with photoexcited polar vibrations{\large$^8$}, photonic crystals
with two-photon transitions{\large$^9$}, and plausible inverted 
superconductors{\large$^{10}$}.

The feasibility of influencing coherent processes by quantum effects opens 
a novel way of regulating the characteristics of collective atomic radiation.


\vskip 2cm

\begin{thebibliography}{99}

\bibitem{1}
L. Mandel and E. Wolf, {\it Optical Coherence and Quantum Optics}, Cambridge
University, Cambridge, 1995.

\bibitem{2}
R.A. Minlos, Ed., {\it Euclidean Quatum Field Theory}, Mir, Moscow, 1978.

\bibitem{3}
V.I. Yukalov, Coherent radiation from polarized matter, {\it Laser Phys.},
{\bf 3}, 870--894, 1993.

\bibitem{4}
V.I. Yukalov, Theory of coherent radiation by spin maser, {\it Laser 
Phys.}, {\bf 5}, 970--992, 1995.

\bibitem{5}
V.I. Yukalov, Nonlinear spin dynamics in nuclear magnets, {\it Phys. Rev. B},
{\bf 53}, 9232--9250, 1996.

\bibitem{6}
V.I. Yukalov, Nonadiabatic dynamics of atoms in nonuniform magnetic fields, 
{\it Phys. Rev. A}, {\bf 56}, 5004--5013, 1997.

\bibitem{7}
V.I. Yukalov and E.P. Yukalova, Cooperative electromagnetic effects, {\it
Phys. Part. Nucl.}, {\bf 31}, 561--602, 2000.

\bibitem{8}
L. Lauck, A.R. Vasconcellos, and R. Luzzi, Nonlinear effects in the 
production of highly photoexcited polar vibrations, {\it Phys. Rev. B}, 
{\bf 46}, 6150--6160, 1992.

\bibitem{9}
A.M. Basharov, Specific features of the resonance interaction of a coherent 
wave with an ensemble of impurity atoms of a photonic crystal, {\it Opt. 
Spectrosc.}, {\bf 90}, 496--500, 2001.

\bibitem{10}
M.I. Kurkin, Can isolator be superconducting? {\it Nature}, N{\bf 4}, 3--9,
2001.

\end{thebibliography}
\end{document}